\documentstyle[prl,aps,epsf]{revtex} \def\narrowtext{} \tighten \twocolumn
\input epsf.sty
\begin{document}
 
\title{Schottky-Enabled Photoemission in a RF Accelerator Photoinjector - Possible Generation of Ultra-Low Transverse Thermal Emittance Electron Beam}
\author{
        Zikri M. Yusof, Manoel E. Conde, and Wei Gai
          }
\address{
         High Energy Physics Division, Argonne National Laboratory, Argonne, IL 60439
             }
\address{%
\begin{minipage}[t]{6.0in}
\begin{abstract}
We present a clear signature of the Schottky effect in a RF photoinjector using photons with energy lower than the Mg cathode work function. This signature is manifested by the shift in the RF phase angle for the onset of the detection of photoelectrons via single-photon absorption and allows for a reasonable estimate of the field enhancement factor. This is a viable method to generate an electron beam with very low thermal emittance and thus, a high brightness beam.
\typeout{polish abstract}
\end{abstract}
\end{minipage}}

\maketitle
\narrowtext

Understanding the factors that influence the production and quality of electron beams in a photoinjector is vital in advancing the technology of free-electron lasers (FEL) and linear colliders. This is especially true in the ability to obtain electron beams with high-brightness, which is a combination of high charge and low emittance characteristics. The latter is often limited by the thermal emittance, $\varepsilon_{th}$, of the photoelectrons, which reflects the distribution of the transverse electron velocities as they emerge from the cathode, and continues to be present in the beam even as the particles are accelerated in the axial direction. This limits how tightly a beam can be focused at the interaction point of a collider and degrades the growth rate of a FEL radiation. We report on the first direct observation of the Schottky-enabled photoemission from a photoinjector cathode due to a RF field using photons with energy lower than the cathode work function. The single-photon photoemission process is possible due to the lowering of the effective work function of a metal by the RF electric field ($E$-field) on the cathode (Schottky effect). This effect can be used to significantly lower $\varepsilon_{th}$ of an electron beam, opening up new possibilities in the quest for high brightness beams.

The minimum transverse emittance is limited by $\varepsilon_{th}$. For a polycrystalline photocathode, $\varepsilon_{th} \propto \sigma\sqrt{E_{k}/mc^{2}}$ and the uncorrelated kinetic energy is $E_{k} = h\nu - \Phi_{0} + b\sqrt{\beta E(\theta)}$, where $h\nu$ is the photon energy, $\Phi_{0}$ is the work function, $b = \sqrt{e/4\pi\epsilon_{0}}$, $\sigma$ is the RMS laser spot size, and $\beta$ is the local field enhancement factor that depends on the cathode surface properties.\cite{Wang,Wang2,Clendenin} Here, $E(\theta)$ is the applied field on the cathode at the injection phase $\theta$. The presence of the $E$-field causes the lowering of the effective work function that is defined as $\Phi_{eff}=\Phi_{0} - b\sqrt{\beta E(\theta)}$. From the above expressions, one can see that $\varepsilon_{th}$ can be reduced by decreasing the beam size. However, this is limited by space-charge effects. Alternatively, $\varepsilon_{th}$ can be reduced by decreasing $E_{k}$. This is seen clearly in Ref. \cite{Clendenin} where the maximum angle of emission, $\phi_{max}$ approaches zero as $\Phi_{eff}\rightarrow h\nu$. We will discuss a method of doing this using the Schottky effect.

Previous indications of the Schottky effect in a photoinjector gun came from the dependence of the amount of charge detected with the RF phase on the cathode.\cite{Wang,Kobayashi,Gibson} Studies of the quantum efficiency (QE) of cathodes also indicate that this parameter changes with applied $E$-field.\cite{Kobayashi,Garcia,Suberlucq} We extend these studies further by using photons with $h\nu < \Phi_{0}$. Ordinarily, single-photon photoemission does not occur under this condition. Here, we show that within the settings of a RF injector, there is a clear onset of photoemission when a certain $E$-field strength is applied to the cathode surface. This technique allows us to make a reasonable estimate of the field enhancement factor.

The Mg cathode (diameter = 2.8 cm) was made from a solid Mg rod. The surface was polished using diamond powder slurry up to 3 $\mu$m grit. A SEM image showed a surface roughness of the order of 1 $\mu$m while an X-ray spectrum indicated a clean Mg surface with no detectable impurities. The cathode was installed in a 1 1/2 cell, 1.3 GHz standing-wave RF gun at the Argonne Wakefield Accelerator facility (Fig. 1). The base pressure in the gun is $\sim 5 \times 10^{-10}$ Torr with an operating pressure of $\sim 8 \times 10^{-10}$ Torr. Photons of $h\nu=3.3$ eV ($\lambda = 372$ nm) were generated with 1 mJ per pulse and a pulse width of 6 to 8 ps FWHM. The laser enters through the input window and is reflected onto the Mg cathode by the aluminum-coated face of a Ce-doped YAG crystal. There is a similar setup at a different input window for the 5 eV ($\lambda = 248$ nm) laser but using a dielectric mirror that is positioned slightly off axis. Photoelectrons produced by the cathode are then accelerated by the RF field. The laser pulse can be injected at various phases of the RF cycle. Varying this injection phase allows us to vary the $E$-field that is applied to the cathode when the photoelectrons are liberated from the surface. The total charge is then detected by an integrating charge transformer (ICT) at the gun exit. The transverse electron beam profile is obtained using the same YAG crystal and is viewed by a camera. By adjusting the fields from a set of solenoids (Fig. 1) and viewing the beam profile, we can ensure that all the generated electrons that leave the gun passed through the ICT. The details of the beamline and laser system can be found in Ref. \cite{Conde}.

\begin{figure}
\vspace{-0.5cm}
\epsfysize=4.9in
\epsfbox{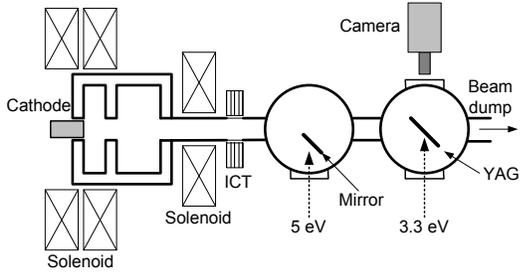}
\vspace{-7.8cm}
\caption{Schematic diagram of the RF photoinjector. There are two separate entrance windows for 5 eV and 3.3 eV laser.}
\label{fig1}
\end{figure}

Fig. 2 shows the measured charge as a function of $\theta$, for various RF amplitudes $E_{max}$. Since $\theta$ is the phase of the RF when the laser pulse hits the cathode, the magnitude of the oscillating $E$-field on the cathode surface when the photoelectrons are emitted is $E(\theta) = -E_{max} \sin(\theta)$. One might expect to detect charge exiting the gun when $0 < \theta < 180^{\circ}$, since at other phases (within a cycle) the electric force would push the electrons back into the cathode. However, within this $180^{\circ}$ range, for $\theta > 130^{\circ}$, even when the electrons could initially leave the cathode, they are unable to exit the gun because the axial electric field switches direction before they escape the gun cavity. Thus, the detection range is only $0 < \theta < 130^{\circ}$. Fig. 2a shows several scans using the 5 eV photons, which are above the Mg work function $\Phi_{0}=3.7$ eV. Although the total charge differs for different $E_{max}$, each curve is qualitatively similar to each other, having a roughly asymmetric bell-shaped profile and approximately the same phase range where the photoelectrons are detected. Previous studies on why such measurements do not yield the expected ``flat-top'' curves pointed to the Schottky effect as the dominant cause.\cite{Wang,Kobayashi,Gibson} While this is certainly plausible, other factors such as space-charge effects and transport issues may also affect such phase scan results. As will be shown, our work detects the Schottky effect in a different and more direct manner. In the process, we discover a viable technique to possibly generate electron beams with low $\epsilon_{th}$.

Figure 2b shows the same measurement done with 3.3 eV photons, which is below $\Phi_0$. The laser spot size impinging on the cathode is $\sim 1$ cm in diameter. Surprisingly, we observe no qualitative differences between these and the ones taken with the higher photon energy in Fig. 2a. There is a drop in the amount of charge detected, but there are no significant differences in the range of phase for the detection of photoelectrons. We attribute the production of these photoelectrons to the two-photon photoemission (2PPE) process as the dominant mechanism, which we verify later. Unfortunately, this process masks any clear signature of the Schottky effect.

To reduce the effects from the 2PPE, we expand the laser spot size from 1 cm to 2 cm in diameter. This reduces the photon density per unit area impinging on the cathode and lowers the occurrence of the 2PPE. A repeat of the measurement produces a stark contrast from before, as can be seen in Fig. 2c. We now detect a change in $\theta$ for the onset of photoemission, shifting to higher values as $E_{max}$ decreases. No shift is detected for the three highest values of $E_{max}$. This is due to a combination of the resolution and accuracy of our detection, and also because the $E$-field changes more rapidly over a smaller phase angle. When there is a shift in the onset $\theta$, the value of $E_{0}$, the $E$-field at the onset of photoemission, is 9.2, 8.5, and 11 MV/m for $E_{max} = $28, 17, and 14 MV/m, respectively. This indicates that $E_{0}$ is a relatively constant value. The $E$-field needs to be at or above this value for photoemission to occur. This is a clear signature of the Schottky effect and the first direct observation of this effect in an RF photoinjector using this technique.

\begin{figure}
\vspace{-0.9cm}
\epsfysize=4.5in
\epsfbox{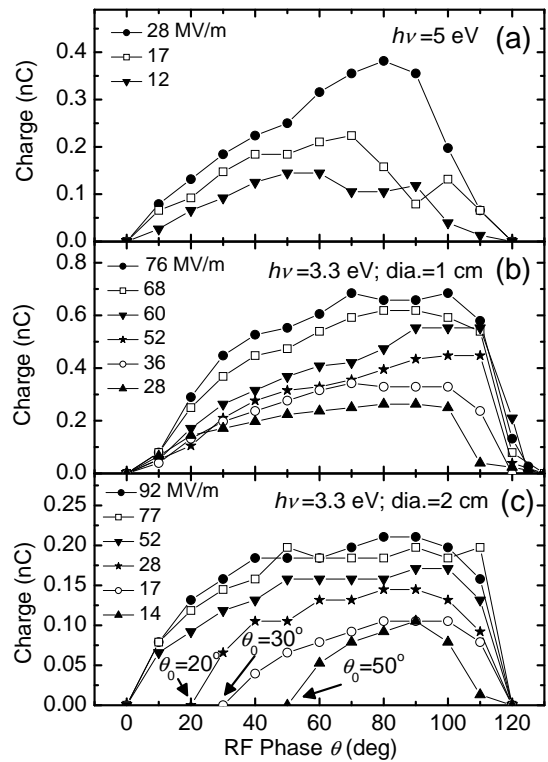}
\vspace{-0.2cm}
\caption{Charge emitted at various RF phase $\theta$. The numbers in the legends are the RF amplitude $E_{max}$ so that $E$-field on the cathode is $E(\theta) = -E_{max}sin(\theta)$. (a) Charge produced by 5 eV photons. (b) Charge produced by 3.3 eV photons with a 1 cm laser diameter. (c) Charge produced by 3.3 eV photons with a 2 cm laser diameter. $\theta_{0}$ is the phase angle ($\pm 5^{\circ}$) at the onset of photoemission.}
\label{fig2}
\end{figure}

In a single-photon emission process, the total photoelectron charge $Q$ can be described by\cite{Wang} $Q \propto (E_{k})^{x}$, where $x$ is not precisely known.\cite{exponent} Our analysis is independent of this exponent since $E_{k}=0$ at the photemission threshold. Using this, we obtain the value of $\beta$ for our cathode to be between 6 and 7. Note that there can be a range of values for $\beta$ at different locations on the cathode surface. The value that is obtained above represents roughly the largest values of $\beta$ since it is calculated at the initiation point of the photoemission process with the lowest $E$-field applied to the cathode. Emitters with smaller $\beta$ values are ``turned on'' in succession as the $E$-field increases.

Fig. 3 shows the charge produced as a function of the laser intensity for the 1 cm and 2 cm laser spot size. The charge emitted from the 1 cm beam clearly shows a non-linear behavior with the laser intensity, while charge from the 2 cm beam appears to be more linear. We verify this by considering that for each laser pulse, the amount of charge emitted can be written as $Q=a_{n}T^{n}$, where $a_{n}$ is a constant coefficient, $T$ is the total laser energy per pulse, and $n$ is the minimum number of photons required to overcome the work function.\cite{Brogle} However, for our experiment, we believe that we have a simultaneous combination of single-photon and 2PPE processes, but in different proportions for different laser spot sizes. Therefore, the emitted charge is $Q = a_{1}T + a_{2}T^{2}$, where $a_{1}$ and $a_{2}$ are the coefficients for the single-photon and two-photon emissions, respectively. The relative magnitude of the two terms will indicate which process is more dominant over the other. Thus, we fit the data in Fig. 3 with a second order polynomial. The coefficients obtained from the 1 cm data show a more dominant 2PPE process. On the other hand, the 2 cm data show that the single-photon process is now the more dominant. This confirms the explanation of the major differences that we observed between Fig. 2b and 2c. When the two-photon process dominates as in Fig. 2b, the Schottky effect has no significant influence on the onset of the photoemission process. Consequently, we detect photoelectrons over roughly the``full'' RF phase range ($\sim 130^{\circ}$). However, when the majority of the photoelectrons detected are due to the single-photon process as in Fig. 2c, then the influence of the applied field on the cathode can be clearly seen via the shift of the onset of the photoemission process. Only when the $E$-field on the cathode is above some value ($E_{0}$) do we detect photoelectrons. This is a very clear manifestation of the Schottky effect.

Previous studies on various clean metal surfaces have shown a non-linear dependence between the amount of photocurrent emitted and the laser intensity.\cite{Papadogiannis} The non-linearity is not due to multi-photon photoemission, but rather to the transient effects of the electronic excitations and  occurs for laser intensities a few orders of magnitude higher than in our work. Furthermore, similar measurements on Mg surfaces confirm that temperature effects play no significant role in the photoemission process within the intensity range of Fig. 3.\cite{Travini} Hence, we can rule out heating and other transient effects as the cause of the non-linearity observed in Fig. 3.

There appears to be two puzzling observations from Fig. 3. First, the charge detected from the same laser intensity is considerably less when the single-photon photoemission dominates than when the 2PPE process dominates. One expects that the first-order single-photon transition would produce considerably more photoelectrons than the second-order process\cite{Brogle} since the cross-section for the single-photon photoemission is at least three orders of magnitude larger than the 2PPE.\cite{Petite} The second puzzling observation is that the intensity measurement was done at the $E$-field strength of 70 MV/m on the cathode. Even without any field enhancement ($\beta=1$), at this $E$-field level, simple Schottky effect calculations show that $\Phi_{eff} < 3.3$ eV for Mg. This means that the single-photon photoemission should dominate both data sets of Fig. 3, and that the 2PPE should be negligible. As can be seen in Fig. 3, the two-photon process dominates for the 1 cm beam size, and it is still present, but less dominant, for the 2 cm beam, contrary to what is expected.

\begin{figure}
\vspace{0.8cm}
\epsfysize=2.5in
\epsfbox{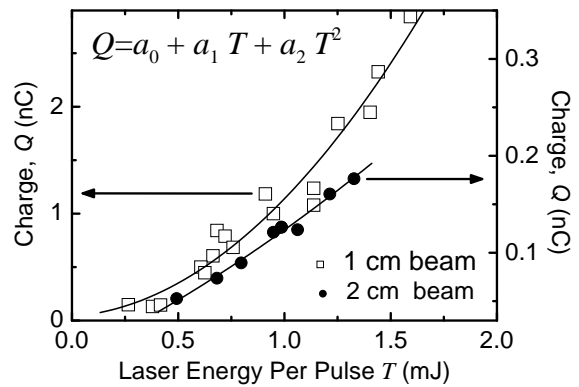}
\vspace{-2cm}
\caption{Charge vs. laser energy per pulse from the 3.3 eV photons. The solid lines are fits to the equation shown. Here, the $E$-field on the cathode is 70 MV/m. For the 1 cm laser spot diameter, the fitting parameters are $a_{0}=0.04, a_{1}=0.14, a_{2}=0.96$. For the 2 cm beam, $a_{0}=-0.01, a_{1}=0.11, a_{2}=0.02$.}
\label{fig3}
\end{figure}

It is highly likely that both puzzling observations were caused by the presence of a layer of MgO on the cathode surface. The Mg cathode was polished and cleaned in air, and remained exposed for about an hour before it was inserted into the photoinjector. Previous X-ray photoemission (XPS) studies on the native oxide layer formed on a Mg surface exposed to air for roughly this duration showed a formation of an oxide layer of between 20 to 30 A in thickness.\cite{Chen} Furthermore, the XPS spectrum clearly revealed that a substantial portion of the photoelectrons collected came from the oxide layer. Thus, for the cathode used in this study, the MgO layer covers the entire cathode surface. Since MgO single crystal has $\Phi_{0}$ of at least 4.2 eV,\cite{Lim} $\Phi_{eff}$ of MgO never drops below the photon energy within the $E$-field range of our study. Hence the 2PPE channel is never eliminated from our measurement, which is why photoelectrons produced via this process are always detected. It means that for the 1 cm beam, the photoelectrons in Fig. 3 could possibly originate from two different processes - the 2PPE from the MgO layer and single-photon photoemission from the Mg surface and regions of $\beta>1$. Secondly, when the beam size is increased to 2 cm, contribution to the photoelectron charge due to the 2PPE process from the MgO layer is no longer dominant since the lower photon density reduces the probability for this transition. The majority of the photoelectrons are now produced from the single-photon photoemission process from Mg. The Mg photoelectrons may come from a combination of possible Mg protrusions due to the surface roughness that were exposed during gun conditioning ($\beta>1$ regions), and from the Mg below the MgO layer.\cite{escape} Considering the small number and area of the protrusions when compared to the MgO surface coverage, and the possibility that the MgO layer may have uneven thickness and may continue to grow even under UHV, this appears to be one plausible explanation on why the detected charge is smaller when the single-photon process dominates.\cite{alternative}

Our technique used in this study has an important application - the possibility to extract electrons with very low $\varepsilon_{th}$. It is well-known that if one could match $\Phi_{0}$ to the photon energy, the photoelectrons are emitted ``cold'' and $\varepsilon_{th}$ is greatly minimized.\cite{Clendenin2} However, in reality, this is not easily achievable because (i) selecting an arbitrary wavelength from a high-powered laser system is not always possible; (ii) cathode surface is typically not ideally smooth and will have a range of $\beta$ that will produce a range of $\Phi_{eff}$; (iii) identical material can have a range of intrinsic $\Phi_{0}$ based on crystallographic orientations;\cite{Lim} and (iv) thermal broadening of the conduction electrons distribution due to the finite cathode temperature will slightly shift $\Phi_{eff}$. The technique that we propose here to produce beams with low $\varepsilon_{th}$ is more realistic and practical.  One only needs to set the photon energy not at one exact, predetermined value, but rather within a range of values below $\Phi_{0}$ of the cathode. Minor imperfections on the cathode surface and variation in the work function of the cathode are less important using this technique. In fact, this technique uses any imperfections resulting in high $\beta$ regions since these will be the regions to photoemit at the threshold. This takes into account any thermal broadening due to the cathode temperature. However, more importantly, we achieve the condition of $h\nu = \Phi_{0}$ or $\Phi_{eff}$ not by tuning the photon energy, but by raising the RF amplitude. This is more realistic for most accelerator photoinjectors. Hence, the technique of using photons with energy below the material's original work function to generate photoelectrons can be used to produce an electron beam with very low $\varepsilon_{th}$.

In conclusion, we have shown the clearest indication of the influence of the Schottky effect on the charge produced in a photoinjector. This is manifested via the clear onset of photoemission above a minimum $E$-field applied to the cathode surface. This is the first ever detection of Schottky-enabled photoemission in a RF photoinjector. Using this effect enables us to make a reasonable estimate of the field enhancement factor on the cathode surface. An important consequence of our study is the possibility of using this technique as a viable means of generating an electron beam with very low thermal emittance. Future plans include a systematic characterization of the emittance of the electron beam generated by this method. We also intend to study surface treatments on the photocathode to further reduce the 2PPE and the field enhancement effects, and to find a more suitable combination of laser energy and photocathode material. This includes using photocathodes with higher QE that would enable the use of a lower power laser.

The authors wish to acknowledge valuable discussions with J.G. Power, A.P. Paulikas, K.-J. Kim, and X.J. Wang, technical assistance from R.S. Konecny, W. Liu, and F. Franchini, and assistance with the SEM system from J.S. Yaeger. This work was supported by ANL LDRD funds and DOE High Energy Physics Division under contract No. W-31-109-ENG-38.


\begin{references}
\bibitem{alternative} We have considered an alternative scenario where all the photoelectrons originated only from the MgO layer - 2PPE from MgO with low $\beta$ while single-photon process from MgO with very high $\beta$. Using similar calculations but with $\Phi_{0}=4.2$ eV, we obtain $\beta \sim$ 50 to 65. Such high enhanced field is not very likely for our cathode surface in this experiment.
\end{references}
\end{document}